\documentclass[sigconf]{aamas}

\usepackage{balance}

\setcopyright{ai4cni}
\acmConference[AAMAS '26]{2nd Workshop on AI for Critical National Infrastructure (AI4CNI-26) @ AAMAS 2026}{May 25 -- 29, 2026}{Paphos, Cyprus}{}
\copyrightyear{2026}
\acmYear{2026}
\acmDOI{}
\acmPrice{}
\acmISBN{}

\acmSubmissionID{<<submission id>>}

\usepackage[T1]{fontenc}

\usepackage{graphicx}
\usepackage{amsmath}
\usepackage{xcolor}
\usepackage{xargs}
\usepackage{ifdraft}
\usepackage{marginnote}
\usepackage{enumitem}
\usepackage{tikz}

\usetikzlibrary{positioning}

\usetikzlibrary{arrows.meta,positioning,fit,calc,shapes.misc, backgrounds}

\newif\ifshowtodos

\newcommandx{\todo}[2][1={}]{%
  \ifshowtodos
    \par\noindent
    \textcolor{red}{\textbf{To do%
      \if\relax\detokenize{#1}\relax\else\ (#1)\fi
      : }}%
    \textcolor{red}{#2}%
    \par
  \fi
}

\definecolor{SlateLight}{HTML}{40403E}
\definecolor{CloudMedium}{HTML}{91918D}
\definecolor{SlateMedium}{HTML}{262625}
\definecolor{IvoryMedium}{HTML}{F0F0EB}
\definecolor{IvoryDark}{HTML}{E5E4DF}
\definecolor{IvoryLight}{HTML}{FAFAF7}
\definecolor{Kraft}{HTML}{D4A27F}

\title[Better Cyber Environments]{Building Better Environments for Autonomous Cyber Defence}

\title[Better Cyber Environments]{Building Better Environments for Autonomous Cyber Defence}

\author{%
Chris Hicks$^{1}$, Elizabeth Bates$^{1}$, Shae McFadden$^{1,2,3}$, Isaac Symes Thompson$^{1}$,\\ [0.5ex]
Myles Foley$^{1}$, Ed Chapman$^{1}$, Nickolas Espinosa Dice$^{4}$, Ankita Samaddar$^{5}$, Joshua Sylvester$^{1}$,\\ [0.5ex]
Himanshu Neema$^{5}$, Nicholas Butts$^{6}$, Nate Foster$^{4}$, Ahmad Ridley$^{7}$, Zoe M$^{1}$, Paul Jones$^{1}$\\ [1ex]
{\normalsize
$^{1}$The Alan Turing Institute,
$^{2}$University College London,
$^{3}$King's College London,\\ 
$^{4}$Cornell University,
$^{5}$Vanderbilt University,
$^{6}$Microsoft,
$^{7}$NSA
}
}

\renewcommand{\shortauthors}{Hicks et al.}

\begin{abstract}
In November 2025, the authors ran a workshop on the topic of \emph{what makes a good reinforcement learning (RL) environment for autonomous cyber defence (ACD)}. This paper details the knowledge shared by participants both during the workshop and shortly afterwards by contributing herein. The workshop participants come from academia, industry, and government, and have extensive hands-on experience designing and working with RL and cyber environments. While there is now a sizeable body of literature describing work in RL for ACD, there is nevertheless a great deal of tradecraft, domain knowledge, and common hazards which are not detailed comprehensively in a single resource. With a specific focus on building better environments to train and evaluate autonomous RL agents in network defence scenarios, including government and critical infrastructure networks, the contributions of this work are twofold: 
(1) a framework for decomposing the interface between RL cyber environments and real systems, and
(2) guidelines on current best practice for RL-based ACD environment development and agent evaluation, based on the key findings from our workshop. 

\end{abstract}

\keywords{Autonomous Cyber Defence, Reinforcement Learning, Resilience}

\begin{document}

\pagestyle{fancy}
\fancyhead{}

\maketitle          

\section{Introduction}

Cyber attacks pose serious risks to individuals, businesses, and governments, whose daily operations all depend on networked systems. In recent history these risks have manifested as costly attacks on critical systems including nuclear facilities~\cite{kushner2013real} and power grids~\cite{geiger2020analysis}. The risk posed by threat actors is managed through a combination of technical controls (e.g., antivirus software, secure boot, air gaps), policy-based measures (e.g., a white list of allowed IP addresses, password policy), and human-focussed measures (e.g., phishing awareness, incident response drills). Despite these preventative measures the asymmetric incentives of adversaries~\cite{anderson2006economics}, and emerging capabilities enabled through AI misuse~\cite{anthropicEspionage2025}, make cyber attacks both a significant and growing problem.

To counter such threats, agent-based Autonomous Cyber Defence (ACD) systems that can immediately monitor, adapt, and respond are key. In particular, Reinforcement Learning (RL), a subset of Machine Learning (ML) that learns through interaction with an environment has received significant interest in the literature ~\cite{han2018reinforcement, kvasov2023simulating, bates2023reward, goel2024optimizing, terranova2024leveraging, foleyCAGEI22, foleyCAGEII22, hicksCAGE3_23}. 
The unique advantage of RL, compared to agents based on pre-trained generative models, is the ability to learn how to achieve a specific goal exclusively from interacting with an environment. RL agents, therefore, do not rely on prior human knowledge or understanding. This is especially important for ACD as adversaries frequently exploit the assumptions made when systems were designed and configured.

For RL to succeed at ACD, there are at least three core challenges: 
(1) the environment must accurately and efficiently represent the real-world network defence problem at an appropriate level of fidelity (e.g., via a cyber environment) including, but not limited to, the network topology, attacker behaviour, and the functionality of software components; 
(2) proper consideration must be given to the interface between the learning agent and the environment, including the identification of a suitable numerical reward mechanism; 
and (3) a robust evaluation methodology, including appropriate measures of success and statistical significance, must be implemented. Towards addressing the core challenges in ACD, this work makes the following main contributions: 
\begin{itemize}[leftmargin=*]
    \item  We \textit{propose a framework} for decomposing the components and modelling choices involved in mapping between ACD environments and real systems.
    \item  We \textit{provide guidelines} on best-practice in relation to each framework component, based on the findings from our workshop, providing a valuable reference for anyone engaged in cyber environment development.
\end{itemize}

The remainder of this paper is organised as follows: Section~\ref{sec:background} provides a background on formulating ACD for RL; Section~\ref{sec:framework} details our framework and how each environment component contributes to the transfer of agent performance from training to deployment; Section~\ref{sec:guidelines} distils the insights gathered from our workshop into best-practice guidelines; Finally, Section~\ref{sec:related} reviews related work and Section~\ref{sec:conclusion} concludes the paper.

\subsubsection*{Workshop Methodology}
This paper is informed by a structured online workshop bringing together 25 domain experts from across academia, industry and government with experience in RL-based cyber defence.

The workshop was organised into three facilitated sessions, each targeting a distinct aspect of ACD environment design: (1) limitations of current environments from a cybersecurity perspective, (2) key modelling, design, and testing considerations, and (3) identification of best practices and prioritisation of open challenges.
Each session utilised the same structure: participants were first presented with a set of guiding questions and given time to independently record their responses on a shared virtual whiteboard (Miro). This was followed by a group discussion where the contributions were reviewed and elaborated upon. Facilitator notes were taken throughout to capture points of agreement, nuance and emphasis that emerged during the discussions. The virtual whiteboard served as both a participation tool and a record, allowing participants to contribute asynchronously and ensuring that the collective output was not limited to verbal contributions.

Following the workshop, the whiteboard contents and facilitator notes were analysed thematically to identify recurring concerns and actionable recommendations. All contributors had additional expertise or references to offer beyond what was captured during the session, they were invited to contribute directly to the manuscript or add comments. The workshop was conducted under Chatham House rules to encourage candid discussion; as such, individual contributions are not attributed in this paper.

\section{RL for Cyber Defence} \label{sec:background}

Both RL and cyber defence are large fields of study comprising many problem types and solution methods. In this work we focus on formulating cyber defence as an RL problem. This section firstly clarifies the definition of an RL problem and then details the scope of network defence commonly encountered in ACD.

\subsection{The RL Learning Problem} \label{RL_Learning_problem}

RL is a field of study, a class of problems, and a class of solution methods for solving such problems~\cite{Sutton2018}. It is the authors' view that RL solution methods (e.g., model-free, on-policy) and algorithms that apply them (e.g., PPO, DQN) are not the main bottlenecks in RL for ACD. Instead, the core challenge of ACD remains how best to formulate a particular network defence problem within the RL framework. 

Informally, the RL learning problem is one where an agent learns how to act in an environment over time to achieve a goal. At each step, the agent selects an action, and then receives an observation and a reward. Through sequential interaction with the environment, the agent learns a policy that aims to maximise the cumulative expected reward. This interaction is modelled as a Markov Decision Process (MDP) with the assumption (a.k.a the Markov property) that the future is independent of the past given the present. In other words, the distribution of the next observation and reward must depend only on the current observation and action (and not those seen or taken previously).

More formally, Sutton and Barto~\cite{Sutton2018} define an RL task; one instance of the RL learning problem, as a complete specification of an environment. An RL task is typically formalised as an MDP $\mathcal{M}$ defined by the tuple:
\[
\mathcal{M} = (S, A, P, R, \gamma)
\]
where $S$ is the set of states, $A$ is the set of actions, $P(s' \mid s, a)$ is the state-transition probability function, $R \subset \mathbb{R}$ is the reward space, and $\gamma \in [0,1]$ is the discount factor. Given the current state $s\in S$ and action $a\in A$, the transition function $P(s' \mid s, a)$ gives the probability of the next state $s^\prime$:
    \[
    p(s' \mid s, a) = \Pr\left(S_{t+1}=s' \mid S_t=s,\; A_t=a\right)
               = \sum_{r \in R} p(s', r \mid s, a)
    \]
\newpage

Given an MDP $\mathcal{M}$, the objective is to learn a policy $\pi(a\mid s)$ that maximises the discounted sum of future rewards $G_t$ after $t$ time steps:
\[G_t = \sum_{k=0}^{\infty} \gamma^k r_{t+k+1}\]
i.e., the optimal policy satisfies $\pi^\star = \arg\max_{\pi} \; \mathbb{E}_{\pi}\left[G_t\right]$. 
The horizon for future rewards is most often bounded by an \textit{episode}, which starts in an initial state and ends upon arriving at a terminal state (or condition). The horizons induced by episodes can be \textit{fixed}, if termination occurs after a fixed number of steps or \textit{variable}, if termination requires certain environmental conditions to occur (e.g., win or loss conditions).   

Finally, when the state space of the environment is not able to be fully observed by the agent, the task is instead modelled as a partially observable MDP (POMDP) defined by the tuple $\mathcal{P} = (\mathcal{S}, \mathcal{A}, P, R, \Omega, O, \gamma)$ where $\mathcal{S}, \mathcal{A}, P, R, \gamma$ are as in an MDP and additionally $\Omega$ is the set of observations and
$O(o \mid s,a)$ is the observation model.

\subsection{Cyber Defence as an RL Task}

Some famous RL tasks (e.g., chess) fit the RL paradigm without significant alteration. However, cyber defence is a complex set of real world problems that does not naturally present a single RL task~\cite{molina2025training}. Beyond comprising multiple tasks and objectives that may change across time, cyber defence is a challenging application for RL because the state of the environment is both extremely large and almost never completely known; adversaries and users introduce non-stationary dynamics; representing cyber security goals as scalar rewards is difficult; operational constraints may be difficult to represent in the action space; and accurate simulators are necessary (e.g., for safety and efficiency) but challenging to build and interface with correctly. Nevertheless, RL offers a principled framework for sequential decision making under uncertainty, enabling agents to learn potentially novel strategies for defending networks in excess of current human-designed approaches. 

\begin{figure}[h]
    \centering
    \includegraphics[width=0.8\linewidth]{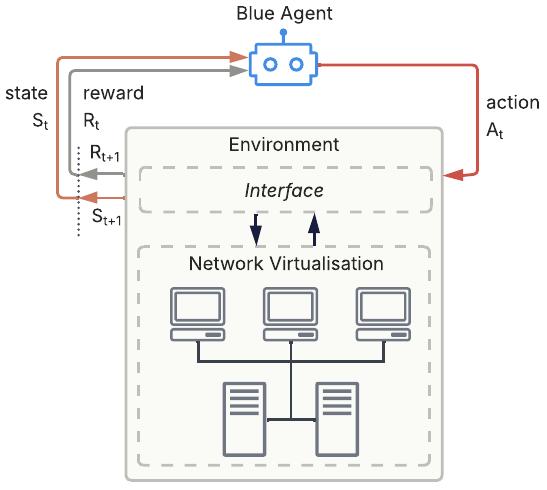}
    \caption{Network defence as an RL task.}
    \label{fig:RL_task_for_ACD}
\end{figure}

Although cyber defence is not a single problem, cyber environments have thus far converged on a relatively constrained interpretation: a defensive blue agent is tasked with defending a network of hosts against one or more fixed-strategy adversaries. 


The action space is usually an enumeration of high-level command and control (C2) techniques (e.g., scan, deny, restore) and the observation space (i.e., the real network state is not completely known) is a vector of selected attributes from the underlying network (e.g., one bit per host to indicate whether a compromise has been detected). The reward function is typically engineered by domain experts and includes penalties and/or incentives for various network states and actions (e.g., -2 when a host is compromised, -1 for restoring a host, +0.1 for making no intervention). In other words, defending the network becomes the task of minimising the number of compromised hosts whilst making conservative use of costly actions. 

Thus, this network defence task of ACD is commonly framed as a finite horizon sequential decision-making problem in which a blue agent receives partial observations from an environment comprising a network of hosts, an adversarial red agent, and possibly benign green agents representing regular user activity. At each time step, the blue agent chooses an action and receives both a new observation, and a reward, based on the state of the environment and which action was chosen. 

Crucially in this framing, the environment comprises everything that is not the blue agent: the network of hosts, regular users, and the adversary, see Figure~\ref{fig:RL_task_for_ACD}. Since the real network is likely of operational significance, and the RL learning framework includes learning from ``mistakes'' to choose better actions over time, a simulator (a.k.a cyber environment) is the standard way to train a network defence agent. Whether blue agents are useful for the real network defence problem, in the real network, therefore depends significantly on how well the cyber environment models the real environment.



\section{A Framework for ACD Environments}\label{sec:framework}

\begin{figure}[h]
\centering
\begin{tikzpicture}[
  font=\sffamily\small,
  line cap=round,
  line join=round,
  ink/.style={draw=Kraft},
  softbox/.style={
    draw=IvoryDark,
    fill=IvoryLight,
    rounded corners=3pt,
    inner xsep=8pt,
    inner ysep=6pt
  }
]

\node[font=\bfseries, text=SlateMedium] (root) at (-2.3,0.95) {Sim-to-Real Gap};

\node[softbox, anchor=north, align=center, text=SlateMedium] (left) at (-4.2,-0.45) {%
  \textbf{Virtualisation Gap}\\[-1pt]
  1.\ Network and Host\\
  \hspace{1.0em}Simulation\\
  2.\ User and Threat\\
  \hspace{1.0em}Simulation
};

\node[softbox, anchor=north, align=center, text=SlateMedium] (right) at (-0.4,-0.45) {%
  \textbf{Modelling Gap}\\[-1pt]
  1.\ Sequence Modelling\\
  2.\ Observation Modelling\\
  3.\ Action Modelling\\
  4.\ Reward Modelling
};

\draw[ink, line width=0.8pt] (root.south) -- ([yshift=4pt]left.north);
\draw[ink, line width=0.8pt] (root.south) -- ([yshift=4pt]right.north);

\end{tikzpicture}
\caption{The breakdown of Sim-to-Real Gap components.}
\label{sim2real_breakdown}
\end{figure}
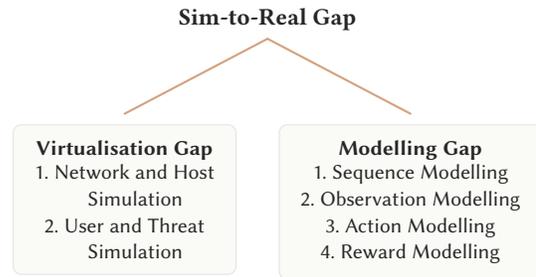

The divergence between environments and real-world systems, known as the \emph{sim-to-real} gap, hinders the transfer of agent performance from training to deployment. Designing high fidelity network environments can help close the gap but is not entirely sufficient. Equally, the formulation of the RL task matters, as this fundamentally constrains what the agent can learn. Therefore, the sim-to-real gap is composed of two key components: the virtualisation of the environment and the modelling of the task, see Figure~\ref{sim2real_breakdown}. These two components capture the whole learning environment as experienced by the agent. Insufficiency in either simulation or modelling limit the real world performance ceiling of RL agents artificially by creating poorly aligned and unnecessarily challenging learning tasks.


\subsection{The Virtualisation Gap}
\begin{figure}[h!]
    \centering
    \includegraphics[width=0.95\linewidth]{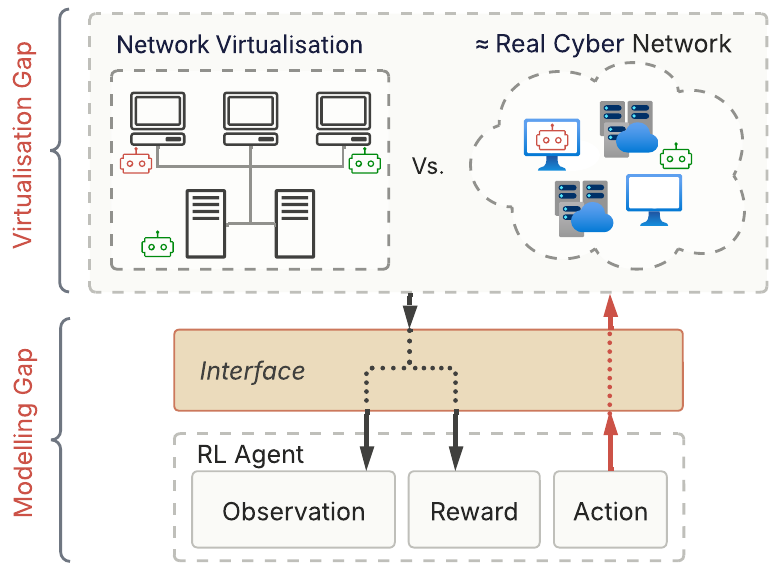}
    \caption{The virtualisation vs. modelling gap taxonomy.}
    \label{fig:sim_gap}
\end{figure}

The virtualisation gap encompasses the difference between the simulated and real-world network defence problems, seen in Figure~\ref{fig:sim_gap}. If RL agents could be trained on a real production network, with real users and attackers, then there would be no virtualisation gap. The virtualisation gap can be split into two components: (1) the network and host simulation including its topology, operational characteristics, and host behaviour, and (2) user and threat simulation.

\begin{description}[topsep=1.5ex, itemsep=1.5ex, leftmargin=0ex, listparindent=2ex]
    
    \item[Network and Host Simulation] 
    High-fidelity simulation of the network and hosts is both feasible, owing to widely available virtualisation software, and preferable for minimising the simulation gap. Low fidelity cyber environments offer high efficiency, making it cheaper to train RL agents, but introduce the significant risk of simulator artifacts. Agents learning from artifacts may perform very well in the simulator yet behave poorly, and unpredictably, in the real world.

    \item[User and Threat Simulation]
    Both regular network users and adversarial threats are an essential part of modelling a network defence problem. Within a cyber environment these are represented by green and red agents, respectively. In proportion to the realism of their behaviour both green and red agents have a large impact on the simulation gap.
    Since red agents model adversarial behaviour, they determine the distribution and severity of threats encountered by the blue agent in the environment. Therefore, if the strategies they simulate are not sufficiently realistic (e.g., simplistic, predictable, or overt) then the defensive policies trained will be brittle with marginal success transferring to real-world deployments. Green agents are of equal importance as without benign user simulation the blue agent may learn policies that succeed at defence but significantly disrupt normal traffic.

\end{description}

\todo[]{Potentially split up figure for the two sections.}

\subsection{The Modelling Gap}
The modelling gap comprises the interface between the network defence problem (e.g., simulation model) and the RL task (i.e., a POMDP), seen in Figure~\ref{fig:sim_gap}. 
The modelling gap includes observation modelling (e.g.,a vector derived asynchronously from network state information), action modelling (e.g., an enumeration of actions and how they are executed on network hosts), sequence modelling (i.e., how POMDP time steps relate to the occurrence of network events) and reward modelling (i.e., how to provide scalar rewards in alignment with real-world goals). Modelling choices significantly impact how effective learned agent policies are, irrespective of the simulation gap, but are also constrained by simulation fidelity. 

\todo[]{Circle back after best practice}


\begin{description}[topsep=1.5ex, itemsep=1.5ex, leftmargin=0ex, listparindent=2ex]

\item[Sequence Modelling.]

Hundreds of processes run concurrently on modern operating systems and network packets are sent, reordered, dropped and received between devices without coordinated clocks. Sequence modelling captures the temporal relationship between continuous, asynchronous network activity and the discrete-time sequence of agent actions, observations and rewards required for RL, seen in Figure~\ref{fig:sequence_modelling}. Sequence modelling is crucial because the action, observation and reward modelling components must preserve causal ordering, allowing meaningful state transitions and reward attribution, which all depend on understanding how the network, hosts and other agents may advance between each time step. The MDP discount factor $\gamma$, which determines an effective prior over the relevance of future rewards, and the RL task episode length when fixed, both typically assume fixed per-step wall time and therefore depend greatly on sequence modelling.


\begin{figure}[h]
    \centering
    \includegraphics[width=1.0\linewidth]{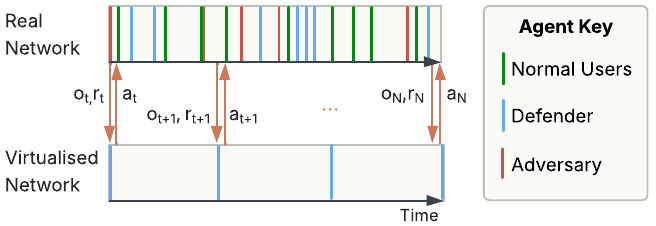}
    \caption{Sequence Modelling: the relationship between continuous, asynchronous real-world network activity (top) and the discrete-time RL sequence trajectory (bottom).}
    \label{fig:sequence_modelling}
\end{figure}

\item[Observation Modelling.]\label{sec:observation_modelling}
The true underlying state of the network is far too large to input to any algorithm (e.g., a single workstation hard disk may have 1TB of state information) and is distributed over many devices. Thus, the network state must be aggregated and parsed to provide vector observations for the RL agent. The crucial modelling constraint of the underlying state is the Markov property: the distribution of the next state and reward can only depend on the current state and action and not those seen or taken previously. Thus, the observation should strive to approximate the Markov property as closely as possible, including any data reasonably available to a defender, whilst respecting realistic monitoring limitations. At the same time, irrelevant features might increase the attack surface and, when featuring high-variance, increase the learning difficulty.

\item[Action Modelling.]
In network defence, operations are usually carried out using highly expressive graphical and command-line user interfaces that do not directly provide the discrete or continuous actions required by RL. Thus, action modelling involves constructing a set of (discrete, continuous, or a combination) of actions that map to a useful set of choices relevant to the network defence task.

\item[Reward Modelling.]
The goal of network defence is to minimise adversarial disruption while maintaining normal operations for non-malicious users. Reward modelling involves identifying scalar incentives and penalties which correspond to achieving one or more network defence goals when their combination is maximised over the discounted horizon. 




\end{description}







\section{Towards Better Cyber Defence Environments} 
\label{sec:guidelines}
This section distils best-practice insights from domain experts into actionable guidelines, designed to maximise RL task effectiveness and real-world transfer, structured around the framework introduced in Section~\ref{sec:framework}. As the main focus of academic work to date~\cite{vyas_survey_26}, this best-practice focusses on single-agent ACD environments for network defence.

\subsection{Characterising the Real-World Problem}
\label{scoping_net_defence_problem}

Before any simulation or RL modelling decisions are made, it is essential to clearly define the target network defence problem. The problem scope will determine which real-world dynamics must be captured, which abstractions may be acceptable, and how problem complexity can be introduced over time. 

\begin{description}[topsep=1.5ex, itemsep=1.5ex, leftmargin=0ex, listparindent=2ex]


\item[Specify the Problem, Constraints \& Success Criteria.]
The first step is to fully specify the network defence problem to be investigated. The specification should characterise the network environment (e.g., topology, scale, host details and roles), operational and business requirements (e.g., user data must have high availability), assets to be protected (e.g., critical hosts, sensitive data), adversarial assumptions (e.g., goals, tactics, techniques, and procedures), and the capabilities and constraints of defensive actors.

\item[Identify Areas of Uncertainty.] A key consideration is characterising uncertainty about the real-world network defence problem. Uncertainty may arise from: the network environment (e.g., variations in network topology and host behaviour), the distribution of expected user and attacker profiles, and defender errors (e.g., malware misclassification). Characterising uncertainty helps to identify generalisation requirements and may improve alignment between the training, evaluation and deployment performance by allowing agents to be robust to unseen distributions of trajectories.

\item[Detail a Minimum Viable Problem.] The combination of a complex full-scale network defence problem, and well-known RL learning reliability challenges~\cite{Agarwal21unstableRL}, is ill-suited to efficient debugging and early experimental iteration. Thus, it is important to construct a well-scoped minimum viable problem that can eventually scale to the required complexity. This abstraction should preserve the core decision-making structure and causal factors of the task while simplifying confounding factors including network scale, state dimensionality, action branching, and non-stationary user and threat behaviour. Once the core problem structure is understood, realism can be systematically reintroduced to recover the full-problem.

\todo[]{potential section on configurability here.}

\end{description}

\subsection{Virtualisation}

Having characterised the minimum viable problem scope and success criteria, the goal is to build a virtual environment that captures the necessary problem structure, users, and adversaries with scope to eventually scale to the full problem. The virtualisation gap constrains the behaviours, uncertainties, and causal relationships that an RL agent can learn from; setting an upper bound on the achievable real-world performance, irrespective of how sophisticated the subsequent RL modelling may be.


\begin{description}[topsep=1.5ex, itemsep=1.5ex, leftmargin=0ex, listparindent=2ex]

\item[Scope Existing Environments.] While a pragmatic first step is to review existing cyber environments~\cite{vyas_survey_26}, any suitably well-characterised (i.e., specific) network defence problem is likely to require significant customisation. If a new environment is judged necessary, then developers must decide between a simulation, emulation, or hybrid approach, balancing efficiency against the risk of simulator artefacts. This decision should be informed by the minimum viable problem, minimising any unnecessary complexity that does not contribute to the learning signal. 

\item[Validate Connection to High-Fidelity.] While this may involve low-fidelity simulation to begin with, it is possible that the correct problem structure and causal factors are not fully understood. Thus, it is critical to validate that the resulting observations, action feasibility, state transitions, and rewards/success criteria correspond to those produced by realistic interactions between emulated or real-world software components, hosts, and network infrastructure. Implementing a bespoke mid-fidelity simulation should be approached with caution as there is a risk of both including unnecessary complexity and failing to match the real-world problem dynamics. 

\item[Inform Design Choices through Problem Scale.] The problem scale should inform the choice of virtualisation technologies, and non-negotiable configurability (e.g., topology, host details, user and threat profiles), ensuring the feasibility of systematic scaling once the minimum viable problem has been satisfied. Finally, the network virtualisation, capabilities and constraints of defensive actors, alongside the measurable criteria for success or failure should inform what monitoring data is available to the defender. The fidelity, noise, and delay characteristics of monitoring information should reflect realistic operational constraints. 
\end{description}

\subsubsection{Network and Host Virtualisation}
At a minimum, the virtualised network should reflect the topology, traffic flows, service dependencies, host roles, and configuration necessary for both defensive and adversarial actions to produce plausible effects. Ideally, high-fidelity virtual machines, containerised instances, and virtual networks running real-world software stacks (e.g., applications, databases and network services) representative of the full-scale problem should be used to provide network and host virtualisation. The network and host state information available for defenders should plausibly represent data produced by the suite of monitoring tools, logs and alerts found in the real-world problem.

\begin{description}[topsep=1.5ex, itemsep=1.5ex, leftmargin=0ex, listparindent=2ex]

\item[Ensure Problem Structure is Preserved.] When a highly abstract network representation is justified by a simple problem structure, the resulting environment dynamics should be validated against a high-fidelity network environment, or using data from the real-world problem, to ensure that problem structure has been preserved as expected. Towards scaling from minimum viable problem to full-scale realism, network topologies, host profiles, services, and monitoring capabilities should be easily configurable. 

\item[Examples of Best-Practice.] Current best-practice examples include Cyberwheel~\cite{OeschCyberWheel2024}, which ensures network and host state information is designed to mimic real-world monitoring and detection pipelines, and NASimEmu~\cite{janisch2023nasimemu} which validates abstract red team simulation dynamics against high-fidelity emulation using a shared simulator-emulator interface.
\todo[]{
\begin{itemize}
\item CSLE from Kim Hammar \citep{Hammar_CSLE}
\end{itemize}
}

\end{description}

\subsubsection{User and Threat Virtualisation}

User and threat virtualisation determine the distribution of benign and adversarial network activity encountered during training and evaluation. Both must be modelled with sufficient realism to ensure agents remain effective in the real-world, avoiding policies that depend on misrepresentative user activity or attacker behaviour.


\begin{description}[topsep=1.5ex, itemsep=1.5ex, leftmargin=0ex, listparindent=2ex]

\item[Move Beyond Fixed Attacker Strategies.] Fixed-strategy, deterministic red agents can be useful when implementing a minimum viable problem, as they remove a source of non-stationarity and thereby ease debugging and early experimental iteration. However, RL policies are known to be brittle under modest shifts in environment dynamics~\cite{PackerGao_generalisable18, samaddar25ood}. As a result, environments that rely solely on stationary adversaries are likely to induce policies which fail to generalise or transfer satisfactorily to real-world settings. Environments should therefore support diverse and configurable red-agent behaviour, including stochasticity, strategy switching, and adaptive decision-making.

Where feasible, environments should support competitive multi-agent training; enabling red and blue agents to co-evolve, and exposing defender policies to diverse offensive strategies. In principle, co-evolution may even lead to state-of-the-art emergent strategies providing pre-emptive mitigation techniques for real-world networks.


\item[Aim for Representative User Activity.]
Legitimate user activity should be simulated with sufficient realism to reflect the diversity and variability of real-world network traffic. Green agent behaviour should generate representative background traffic, service interactions, and benign anomalies that meaningfully interfere with detection and response. Where possible, real-world logs or statistically grounded models of network activity should be used to inform and validate virtual user behaviour, rather than relying on simplistic or scripted profiles.

\item[Consider Generative Green and Red Agents.] 
Virtualisation environments should support multiple approaches to modelling agents. Alongside fixed or learning-based agents, emerging LLM-based, agentic adversaries may allow generating diverse, adaptive, goal-directed agent behaviour without requiring explicit modelling.

\todo[]{Add any best practice citations that are relevant}

\end{description}

\subsection{RL Modelling}

Given a simulated network defence problem, the next step is to develop a suitable RL task formed from observations, actions, temporal structure, and rewards. These modelling choices determine what information the RL agent can learn from, how it can interact, how causality unfolds over time, and how success is incentivised.

\subsubsection{Sequence Modelling}\label{subsubsec:sequence_modelling}
When building an RL task for a network defence problem, sequence modelling ensures that asynchronous real-world network behaviour is properly mapped onto the discrete-time sequence of actions, observations, and rewards that structure RL learning. 

Addressing sequence modelling remains an open challenge for RL in cyber defence, particularly given long-horizon dependencies, asynchronous dynamics, and persistent adversaries. However, explicitly reasoning about temporal structure, action duration, and concurrency is essential for closing the modelling gap and producing agents whose behaviour transfers to the real-world.



\begin{description}[topsep=1.5ex, itemsep=1.5ex, leftmargin=0ex, listparindent=2ex]

    \item[Justify Episodic Structure and Horizon Length.] 
    
    Many ACD environments divide agent-environment interactions into discrete episodes, either with a fixed number of time steps (i.e., fixed-horizon) or explicit terminal states (i.e., episodic). An episodic task formulation simplifies training and evaluation by ensuring that returns are finite and well-scaled, and also by providing a reset mechanism which can increase exposure to important states~\cite{Sutton2018}. However, episodic formulations impose strong assumptions about temporal structure that often do not hold in operational networks. Real-world networks are typically persistent and non-stationary, challenging the possibility of a ``reset'' and discrete episodes. Another difficulty is that episodic formulations impose a bounded decision horizon, eliminating the possibility of learning causal structure beyond the episode bounds and biasing RL agents towards short-term wins.

    
    Best practice therefore is to explicitly justify the temporal task modelling, based on the target problem, rather than adopting an episodic, fixed-horizon by default. If the problem is inherently continuous or long horizon, exemplified by Advanced Persistent Threat (APT) actors, who may gain access to a network months or years before executing their attack, then open-ended or infinite-horizon formulations should be considered. For example, Hammar and Stadler~\citep{Hammar_Intrusion_Prevention_through} formulate network defence as an optimal stopping problem, where the defensive agent has two possible actions: ``stop'', corresponding to a defensive intervention, and ``continue''. 

    \item[Account for Temporal Variability and Action Duration.]\hspace{1ex}\\In real networks, defensive actions are neither instantaneous nor uniform in duration. Scanning, isolating hosts, restoring services, and reconfiguring access controls may take seconds, minutes, or longer, and can overlap with other defensive or adversarial activities. Similarly, attacks unfold asynchronously, driven by independent processes and external timing constraints. In contrast, the standard RL assumption is that each time step corresponds to a fixed quantity of wall-time (i.e., each time step is weighted equally when calculating reward attributions).

    Thus, to support real-world transfer, it is best practice to represent temporal variability and action duration explicitly within the environment. Where possible, environments should support variable action durations, varying delays between action initiation and effect observation, overlapping and concurrent execution, and events occurring independently of agent actions (e.g., scheduled system processes). R3ACE~\cite{chapman2025r3ace} implements continuous-time, event-driven RL modelling which is aligned with best practice, although only demonstrated on a very small problem space.


    \item[Reconsider Turn-Based Modelling.]
    Many cyber environments model red, green, and blue agent interactions in a turn-based manner, implicitly enforcing an ordering of each agent's action. This abstraction is not representative of real networks, where attackers, users, and defenders act concurrently, opportunistically, and without coordination.

    At baseline, it is important to be aware that turn-based agent interaction causes non-trivial impacts on RL learning outcomes in ACD~\cite{bates2026rewardsreinforcementlearningcyber, bates2025less}. Ideally, turn-based modelling should be avoided except where the problem structure imparts an ordering distribution that can be well-determined. If turn-based structure is retained for tractability, then the sensitivity of trained policies should be evaluated.


\end{description}

\subsubsection{Observation Modelling}\label{subsubsec:observation_modelling}

Observations in ACD environments are often fixed-size, discrete vectors representing evidence about network status. However, workshop participants repeatedly highlighted that unrealistic ``magic'' observations and overly lossy encodings (e.g., minimal bit-vectors) undermine both realism and transfer, while also obscuring what a defender is assumed to know at decision time.

\begin{description}[topsep=1.5ex, itemsep=1.5ex, leftmargin=0ex, listparindent=2ex]

\item[Define a Realistic Observation Pipeline.]
The information contained in observations should reflect what is genuinely available in practice from realistic sources such as monitoring tools, logs, and alerts. When observations are drawn directly from the raw simulator or emulator state, they risk incorporating features that could not be reliably produced in real deployments (e.g., listing if a host is compromised in the state). This creates a gap between the environment observed by the agent and real-world conditions, as these ``magic'' observations embed convenient but impossible information that undermine practical transferability.
Equally, modelling observations directly on virtualised network state also risk containing less information than a cyber operator would typically have access to, under-representing true monitoring capability. Since magic observations cannot be reproduced outside of the virtualised environment, observation modelling must be grounded in practically available information. For example, the Cyberwheel~\citep{OeschCyberWheel2024} environment demonstrates this by constructing observations directly from existing cyber detection tools.

\item[Consider Observation Representation.] For ACD with a single blue agent, the most common observation design is to aggregate information from all network nodes into a single observation vector, mimicking the design of most traditional RL environments. 
However, this approach can lead to limitations in expressive power and generalisability when using standard neural network architectures~\citep{mern2020exchangeableinputrepresentationsreinforcement}. This is especially problematic in environments with varying network topologies, which can correspond to changing the dimensions of the observation vector. An alternative is to factorise observations into individual network node features, and apply GNN \citep{janisch2023nasimemu,nyberg2024structuralgeneralizationautonomouscyber,dudman2025generalisablecyberdefenceagent, king2025automatedcyberdefensegeneralizable} or attention-based \citep{mern2021autonomousattackmitigationindustrial, symesthompson2025entitybasedreinforcementlearningautonomous} architectures, leveraging their permutation equivariance to produce policies that are more robust across varying network configurations.

\end{description}

\subsubsection{Action Modelling}

To enable learning effective strategies, agents need useful actions that correspond to real-world capabilities. Action modelling requires carefully mapping relevant capabilities into a set of actions that can be chosen from by the RL agent.

\begin{description}[topsep=1.5ex, itemsep=1.5ex, leftmargin=0ex, listparindent=2ex]

\item[Provide a Sufficient and Representative Action Space.] The action space fundamentally defines an agent's capabilities and must therefore provide sufficient scope to enable learning diverse and effective strategies. If the action space is very narrow or abstract the agent will likely be constrained to a small set of behaviours, limiting its capabilities below what may be effective in deployment. A representative action space should map to realistic operations that cyber defenders can perform on the network, preserving the temporal and causal relationships between system state, actions, and their outcomes in the environment.

\item[Consider the Granularity-Capability Trade-off.] Fine-grained actions can lead to large action spaces, increasing the difficulty of learning an effective policy. Whereas coarse abstractions (e.g. Mitre D3FEND~\cite{Mitre} or Open C2~\cite{openC2}) risk removing decision-making capability and breaking causal realism. 
The granularity of actions should reflect the intended operator role and specified network defence problem. Where possible, the granularity should be validated against real-world operational practice and be mapped to actual behaviour in high-fidelity emulation. 
Iteratively increasing the action granularity is consistent with the best practice of starting with an MVP and systematically scaling up to the full network defence problem. As realism scales up, and the action space becomes more fine grained, additional capability and flexibility granted to the blue agent may come at the expense of learning efficiency. This trade-off could be managed through iterations of task modelling. To deal with large combinatorial action spaces, first consider approaches from the RL action space decomposition literature~\citep{MilesEtAl2024RL_ARCD}.
\item[Mask Invalid Actions.] Not every action is necessarily meaningful or feasible in a given state, as determined by the specific network conditions and privileges given to the agent. Allowing agents to select invalid actions can introduce undefined behaviour, and wastes training time on zero-value choices, rather than learning effective strategies. Masking invalid actions ensures the action space appropriately reflects operational reality and supports more efficient learning. Molina-Markham \emph{et al.}~\citep{molina2025training} suggest using the Planning Domain Definition Language (PDDL) for modelling network defence tasks, including using action preconditions to determine masks.
\item[Ensure Effects are Reflected in Observations.] Valid actions that do not have observable consequences do not provide a useful learning signal and introduce noise into training that can impede policy learning. Furthermore, the success of a chosen action is not guaranteed in realistic deployments. Therefore, if available in practice, actions should have a discernible effect that can be observed by the agent, either directly or indirectly over subsequent states. This dependency between action and state modelling should be validated to ensure environments facilitate effective agent training.

\item[Make Configuration Easy.] The interface should facilitate the addition, removal, or changing of constraints on actions to allow for the modelling of specific network defence subtasks. This flexibility enables experimentation with different capability assumptions, promoting the reuse of principled environments and improving reproducibility.

\end{description}

\subsubsection{Rewards Modelling}
The reward function fundamentally defines the learning objective on an RL agent, making its modelling highly consequential when producing ACD agents. An incorrectly specified reward function can lead to policies that appear successful in maximising long-term rewards, but fail to genuinely achieve defensive objectives.

\begin{description}[topsep=1.5ex, itemsep=1.5ex, leftmargin=0ex, listparindent=2ex]


\item[Motivate the Reward Function.] Reward functions should be motivated and justified with respect to the underlying defensive objective. The general goal in network defence is typically the sustained operation of a system that continues to meet user demands while hosts remain uncompromised. The decisions relating to reward function modelling must be justified and documented, ensuring that the chosen reward function reflects operational requirements rather than arbitrary design choices.

\item[Align with the ACD Problem Goal.] It is important to consider the alignment of rewards to the overarching goal. Reward functions fundamentally assign relative value to states and actions, which can unintentionally introduce equivalencies between meaningfully different network states (e.g., restoring a user host versus an operational server). Reward misalignment can lead the agent to ``game'' the task by maximising rewards without producing a valid defensive policy, referred to as reward hacking. Therefore, principled reward design is necessary to avoid reward hacking and produce operationally valuable behaviour.

\item[Simplify Instead of Over Engineering.] Where possible, reward functions should minimise unnecessary shaping and avoid encoding detailed domain heuristics directly into scalar rewards. Highly engineered dense rewards can introduce unintended biases and impose arbitrary trade-offs between defensive actions and network states. As a result, reward functions should ideally be sparse such that only a few, but feasibly reachable, goal-aligned state-action pairs provide a reward signal~\cite{bates2026rewardsreinforcementlearningcyber, bates2025less}. This is achievable in ACD RL tasks where episodes begin in a goal state (i.e., an uncompromised network). Sparse rewards place fewer constraints on agent behaviour, support long-horizon planning, and reduce the likelihood of learning policies that exploit artefacts of reward design rather than achieving goal objectives. However, sparsity must be balanced against exploration and learning stability~\cite{bates2023rewardshapinghappierautonomous}. Scaling from smaller, less complex networks to large networks of interest throughout training, as a form of curriculum learning, may help to mitigate some of these problems.

\end{description}

\subsection{Evaluation}

Comprehensive evaluation methods are critical for building a good RL cyber environment. 
Thorough evaluations in complex RL environments illuminate any unknowns about true agent performance, revealing issues in both modelling and network virtualisation that might otherwise be missed.
The process of realistically simulating or emulating networks is inherently noisy and stochastic. Therefore, robust evaluation measures are needed to give assurances about the reproducibility, statistical significance, and risk profile of agent performance relative to the specified ACD problem. 


\begin{description}[topsep=1.5ex, itemsep=1.5ex, leftmargin=0ex, listparindent=2ex]

\item[Understanding the Learnt Policy.] Typical RL evaluation practices for ACD problems include reporting average episodic rewards~\cite{cage2_kiely2023autonomousagentscyberdefence, primaite, msft:cyberbattlesim}, and variance metrics~\cite{Hammar_Intrusion_Prevention_through, janisch2023nasimemu}. 
Although, this alone is not sufficient to get a complete picture of an agent's learnt policy, nor is it an independent measure of how correct an MDP is. 

\item[Ensure Statistical Validity.] 
%

Sufficient analysis of variance in policy performance is an aspect of evaluation frequently missing in research applying RL to cybersecurity tasks~\cite{mcfadden2026sok}.
Statistical validity is only achievable using multiple independent training runs with different random seeds to establish confidence intervals and ensure reproducibility~\cite{patterson2024empirical}.

\item[Evaluate Behaviour at the System-Level.]  An RL agent will always seek to maximise the reward signal it receives and simply evaluating using average episodic rewards only shows how well the agent has optimised its behaviour with respect to \textit{that} function. 

Hence, evaluating using average episodic rewards is not sufficient to get a full picture of the trained agent's policy, nor is it an independent measure of how correct your task modelling is, reflecting Goodhart’s law: when a metric becomes the optimisation target, it ceases to be a reliable measure~\citep{ashton2021causalcampbellgoodhartslawreinforcement, karwowski2023goodhartslawreinforcementlearning}.

Agent evaluation must move beyond only episodic rewards and extend its metrics to measure system-level activity relevant to the goal: how often are hosts attacked? Which hosts? What kind of attack? Analysis of the blue agent's learnt policy can shed light on where in the network the blue agent is acting most and what the average action distribution looks like. 
This builds a bigger picture of the defensive strategy the agent has learnt and provides granular policy information to critically assess how well the observations, actions and rewards are being modelled.
The information to track these metrics and extract agent trajectories is found at the simulation level of the cyber environment, where the state of the network can be assessed in relation to the foundational network defence goals independently of the RL task.
\end{description}

\section{Related Work}\label{sec:related}
\begin{description}[nosep, leftmargin=0ex, listparindent=2ex]
    \item[Motivations and Limitations of existing ACD Environments.]
    An early effort at designing an RL environment for ACD was FARLAND \citep{molinamarkham2021networkenvironmentdesignautonomous}, ``\textit{a framework for advanced Reinforcement Learning for autonomous network defense}''. This work emphasised the importance of configurability in the underlying environment and the ability to perform curriculum learning by training across different scenarios of varying complexity. FARLAND also incorporated adversarial red behaviour that manipulates the observations of the blue agent, as opposed to simply attempting to infiltrate the network. 
    
    CybORG is a simulator developed for the ``Cyber Autonomy Gym for Experimentation'' (CAGE) challenges \citep{cage_1_cyborg_acd_2021, cage2_kiely2023autonomousagentscyberdefence, cage3_code, cage_4_dev_kiely2025cage, cage_4_writeup_kiely2025exploring}, which has become a standard benchmark environment for the development of RL agents for ACD. The CybORG version most commonly used in the literature is CAGE 2, but a number of authors identify fundamental issues with this environment. CybORG++ \citep{emerson2024cyborgenhancedgymdevelopment} is an iteration on the CAGE~2 version of CybORG that identifies and fixes problems with the original version and offers a lightweight alternative ``MiniCAGE''.

    A more recent environment is Cyberwheel~\citep{OeschCyberWheel2024}, which provides both a simulator and emulator. As part of the motivation for developing Cyberwheel, the authors discuss their attempt to extend the CAGE~2 version of CybORG and identify six core issues that led them to develop a new environment: Lack of network topology configurability, inadequate red agent behaviour, no emulation environment, no visualisation tools, poorly scalable observation design (see section~\ref{subsubsec:observation_modelling}), and dead code. 

    The UK's Dstl have supported the development of various ACD environments at different levels of fidelity, including support of CybORG. As part of Dstl's ARCD (Autonomous Resilient Cyber Defence) programme, four simulators were developed~\citep{MilesEtAl2024RL_ARCD,Short2025EssentialRoleMSAI}. Yawning Titan~\citep{andrew_Yawning_Titan} is a low-fidelity, abstract simulated environment, intended to encourage fast experimental iteration of defensive agent architectures. Within the constraints of its abstract setting, it allows for a great degree of configuration of network topology, node types, reward functions and action types. PrimAITE~\citep{primaite} is a higher fidelity simulated environment allowing for the configuration of Information Exchange Requirements (IERs) and red and green patterns-of-life at a node level (as opposed to simulated red and green agents). In principle, this could allow for training a defensive agent on a large distribution of possible scenarios, provided it is feasible to generate sufficient numbers of realistic pattern-of-life configuration files. Imaginary Yak is a closed-source emulated environment that can operate on containers and virtual machines, and PalisAIDE the highest fidelity environment from ARCD, operating on virtual machines and hardware. 
    One aim of this suite of environments was to transfer an agent trained in a simulated environment (PrimAITE) to a real system (PalisAIDE). 
    Short \citep{Short2025EssentialRoleMSAI} reports there was some success here, but involved a significant engineering challenge, addressing various sim-to-real pitfalls. 
    
    \item[ACD Survey Papers.] 
    Vyas \emph{et al.} \citep{vyas_survey_26} provide a survey of environments for, and approaches to autonomous cyber network defence. Whilst similar in motivation, our paper differs in that we take steps towards formulating a comprehensive framework for building a realistic cyber defence simulator, based on feedback from practitioners during the workshop. Palmer \emph{et al.} \citep{palmer2024deepreinforcementlearningautonomous} provide another review of deep RL for ACD, with a particular emphasis on scalability challenges and an overview of existing ACD environments. In addition to reviewing existing DRL approaches, they outline desirable properties for benchmarking environments that range from appropriate fidelity to the need for emulators.
    

    \item[Multi-agent Reinforcement Learning for ACD.]
    While this paper focuses on single-agent formulations, multi-agent reinforcement learning (MARL) offers a complementary perspective for ACD. 
    In competitive settings, red and blue agents can co-evolve through self-play, exposing defender policies to diverse and adaptive offensive strategies rather than relying on scripted attacker behaviour~\cite{zhang2024survey}. 
    Multiple defensive agents can learn to coordinate across network segments, potentially enabling more scalable and resilient defence architectures. 

    However, multi-agent training introduces additional challenges. Gronauer \emph{et al.} (2021) highlight the issue of non-stationarity due to there being multiple agents acting in a single environment. In MARL, the environment dynamics are non-stationary from the perspective of each individual learner and introduce a moving target problem~\cite{gronauer2022multi}.

    Computational complexity is a further challenge, especially as the joint state-action space increases exponentially in the number of agents~\cite{qu2022scalable, huh2023multi}.
   Despite these difficulties, MARL remains a promising direction for ACD research. Many of the principles discussed in this paper remain relevant in multi-agent settings and extending the framework to such formulations is a valuable direction for future work.

    \item[RL for Network Offence.]
    Whilst this paper focuses on RL for network defence, there is also a large body of research~\citep{yang2025behaviour, hu2020automated, chen2023gail, maeda2021automating, gangupantulu2022using, li2023innes} and RL environments for penetration testing. For example, CyberBattleSim \citep{msft:cyberbattlesim} and NASim \citep{schwartz2019nasim}. NASimEmu \citep{janisch2023nasimemu} extends NASim to have an emulation component, enabling the transfer of agents trained in simulation to an emulated network. 

    \item[RL for Cybersecurity.] Outside of network security, RL has been applied to a range of different cybersecurity applications. McFadden \emph{et al.}~\citep{mcfadden2026sok} surveyed and systematised common pitfalls of the domain at large. Common applications include: 
    \textit{evading detection}~\citep{tsingenopoulos2024train, chen2024llm, hore2025deep},
    \textit{detecting malicious activity}~\citep{praveena2022optimal, McFadden2026DRMD},
    \textit{vulnerability discovery}~\citep{foley2025apirl, al2023sqirl, mcfadden2024wendigo}, blockchain security~\cite{hou2019squirrl, de2024guideenricher}, and hardware security~\cite{gohil2022attrition, luo2023autocat}.
    

\end{description}

\section{Conclusion}
\label{sec:conclusion}

Building effective RL environments for ACD requires principled decisions at every stage of design, from characterising the real-world problem to virtualisation of the network and modelling the agent interface. This paper provides a conceptual framework that maps the core components of ACD environments to two parts of the \textit{sim-to-real} gap, virtualisation and modelling. Using this framework, the paper provides a systematisation of best practice guidelines. 
The central theme of these guidelines is that utility of an ACD environment is determined by how faithfully its design reflects the structure and constraints of the real-world problem; without this grounding, agents risk being optimised for performance in the training environment that does not transfer to capabilities in the operational task.
We hope that the framework and best practice guidelines provided in this paper enable the development of more effective ACD environments. 

\begin{acks}
We would like to additionally thank Dr. Andres Molina-Markham for their contributions to this work both in the initial workshop and during the write-up.
\end{acks}

\bibliographystyle{splncs04}
\bibliography{references}

@inproceedings{hicksCAGE3_23,
  author = {Hicks, C. and Mavroudis, V. and Foley, M. and Davies, T. and Highnam, K. and Watson, T.},
  title = {{Canaries and Whistles: Resilient Drone Communication Networks with (or without) Deep Reinforcement Learning}},
  year = {2023},
  
  
  
  booktitle = {Proceedings of the 16th ACM Workshop on Artificial Intelligence and Security},
  
  
  series = {AISec '23},
}

@inproceedings{foleyCAGEII22,
  title = {{Inroads into Autonomous Network Defence using Explained Reinforcement Learning}},
  year = {2022},
  author = {Foley, M. and Wang, M. and M, Z. and Hicks, C. and Mavroudis, V.},
  booktitle = {Conference on Applied Machine Learning in Information Security (CAMLIS)},
}

@inproceedings{foleyCAGEI22,
  author = {Foley, M. and Hicks, C. and Highnam, K. and Mavroudis, V.},
  title = {{Autonomous Network Defence Using Reinforcement Learning}},
  year = {2022},
  
  url = {https://doi.org/10.1145/3488932.3527286},
  booktitle = {Proceedings of the 2022 ACM on Asia Conference on Computer and Communications Security},
  
  series = {ASIA CCS '22},
  
}

@inproceedings{mcfadden2024wendigo,
  title={Wendigo: Deep Reinforcement Learning for Denial-of-Service Query Discovery in GraphQL},
  author={McFadden, S. and Maugeri, M. and Hicks, C. and Mavroudis, V. and Pierazzi, F.},
  booktitle={IEEE Workshop on Deep Learning Security and Privacy (DLSP)},
  year={2024}
}

@inproceedings{tsingenopoulos2024train,
  title={How to Train your Antivirus: RL-based Hardening through the Problem Space},
  author={Tsingenopoulos, I. and Cortellazzi, J. and Bosansk{\`y}, Branislav and Aonzo, S. and Preuveneers, D. and Joosen, W. and Pierazzi, F. and Cavallaro, L.},
  booktitle={Proceedings of the 27th International Symposium on Research in Attacks, Intrusions and Defenses},
  year={2024}
}

@article{PPO,
  author       = {J. Schulman and
                  F. Wolski and
                  P. Dhariwal and
                  A. Radford and
                  O. Klimov},
  title        = {Proximal Policy Optimization Algorithms},
  journal      = {CoRR},
  
  year         = {2017},
  url          = {http://arxiv.org/abs/1707.06347},
}

@article{chen2024llm,
  title={When LLM Meets DRL: Advancing Jailbreaking Efficiency via DRL-guided Search},
  author={Chen, X. and Nie, Y. and Guo, W. and Zhang, X.},
  journal={Advances in Neural Information Processing Systems},
  
  
  year={2024}
}

@inproceedings{de2024guideenricher,
  title={$\{$GuideEnricher$\}$: Protecting the Anonymity of Ethereum Mixing Service Users with Deep Reinforcement Learning},
  author={De Silva, R. and Guo, W. and Ruaro, N. and Grishchenko, I. and Kruegel, C. and Vigna, G.},
  booktitle={33rd USENIX Security Symposium (USENIX Security 24)},
  year={2024}
}

@article{hou2019squirrl,
  title={SquirRL: Automating attack analysis on blockchain incentive mechanisms with deep reinforcement learning},
  author={Hou, C. and Zhou, M. and Ji, Y. and Daian, P. and Tramer, F. and Fanti, G. and Juels, A.},
  journal={arXiv preprint arXiv:1912.01798},
  year={2019}
}

@inproceedings{gohil2022attrition,
  title={Attrition: Attacking static hardware trojan detection techniques using reinforcement learning},
  author={Gohil, V. and Guo, H. and Patnaik, S. and Rajendran, J.},
  booktitle={Proceedings of the 2022 ACM SIGSAC conference on computer and communications security},
  
  year={2022}
}

@inproceedings{luo2023autocat,
  title={Autocat: Reinforcement learning for automated exploration of cache-timing attacks},
  author={Luo, M. and Xiong, W. and Lee, G. and Li, Y. and Yang, X. and Zhang, A. and Tian, Y. and Lee, Hsien-Hsin S and Suh, G Edward},
  booktitle={2023 IEEE International Symposium on High-Performance Computer Architecture (HPCA)},
  
  year={2023},
  organization={IEEE}
}

@inproceedings{al2023sqirl,
  title={{SQIRL}: Grey-Box Detection of SQL Injection Vulnerabilities Using Reinforcement Learning},
  author={Al Wahaibi, S. and Foley, M. and Maffeis, S.},
  booktitle={32nd USENIX Security Symposium (USENIX Security 23)},
  
  year={2023}
}

@article{praveena2022optimal,
  title={Optimal Deep Reinforcement Learning for Intrusion Detection in UAVs.},
  author={Praveena, V. and V., A and Chinnasamy, P and Ali, I. and Alroobaea, R. and Alyahyan, S. Y. and Raza, Muhammad Ahsan},
  journal={Computers, Materials \& Continua},
  
  
  year={2022}
}

@inproceedings{han2018reinforcement,
  title={Reinforcement learning for autonomous defence in software-defined networking},
  author={Han, Y. and Rubinstein, B. I. and Abraham, T. and Alpcan, T. and De Vel, O. and Erfani, S. and Hubczenko, D. and Leckie, C. and Montague, P.},
  booktitle={International conference on decision and game theory for security},
  
  year={2018},
  organization={Springer}
}

@inproceedings{terranova2024leveraging,
  title={Leveraging deep reinforcement learning for cyber-attack paths prediction: Formulation, generalization, and evaluation},
  author={Terranova, F. and Lahmadi, A. and Chrisment, I.},
  booktitle={Proceedings of the 27th International Symposium on Research in Attacks, Intrusions and Defenses},
  
  year={2024}
}

@inproceedings{foley2025apirl,
  title={APIRL: Deep Reinforcement Learning for REST API Fuzzing},
  author={Foley, M. and Maffeis, S.},
  booktitle={Proceedings of the AAAI Conference on Artificial Intelligence},
  
  
  
  year={2025}
}

@inproceedings{kvasov2023simulating,
  title={Simulating Deception for Web Applications Using Reinforcement Learning},
  author={Kvasov, A. and Sahin, M. and Hebert, C. and De Oliveira, Anderson Santana},
  booktitle={European Symposium on Research in Computer Security},
  
  year={2023},
  organization={Springer}
}

@inproceedings{goel2024optimizing,
  title={Optimizing cyber defense in dynamic active directories through reinforcement learning},
  author={Goel, D. and Moore, K. and Guo, M. and Wang, D. and Kim, M. and Camtepe, S.},
  booktitle={European Symposium on Research in Computer Security},
  
  year={2024},
  organization={Springer}
}

@inproceedings{bates2023reward,
  title={Reward shaping for happier autonomous cyber security agents},
  author={Bates, E. and Mavroudis, V. and Hicks, C.},
  booktitle={Proceedings of the 16th ACM Workshop on Artificial Intelligence and Security},
  
  year={2023}
}

@inproceedings{McFadden2026DRMD,
  title = {DRMD: Deep Reinforcement Learning for Malware Detection under Concept Drift},
  author = {McFadden, S. and Foley, M. and D'Onghia, M. and Hicks, C. and Mavroudis, V. and Paoletti, N. and Pierazzi, F.},
  booktitle = {Proc. of the {AAAI} Conference on Artificial Intelligence},
  year = {2026},
}

@article{yang2025behaviour,
  title={Behaviour-diverse automatic penetration testing: a coverage-based deep reinforcement learning approach},
  author={Yang, Y. and Chen, L. and Liu, S. and Wang, L. and Fu, H. and Liu, X. and Chen, Z.},
  journal={Frontiers of Computer Science},
  
  
  
  year={2025},
  
}

@inproceedings{hu2020automated,
  title={Automated penetration testing using deep reinforcement learning},
  author={Hu, Z. and Beuran, R. and Tan, Y.},
  booktitle={2020 IEEE European Symposium on Security and Privacy Workshops (EuroS\&PW)},
  
  year={2020},
  organization={IEEE}
}

@article{hore2025deep,
  title={Deep packgen: A deep reinforcement learning framework for adversarial network packet generation},
  author={Hore, S. and Ghadermazi, J. and Paudel, D. and Shah, A. and Das, T. and Bastian, N.},
  journal={ACM Transactions on Privacy and Security},
  
  
  
  year={2025},
  
}

@article{chen2023gail,
  title={{GAIL-PT}: An intelligent penetration testing framework with generative adversarial imitation learning},
  author={Jinyin Chen and Shulong Hu and Haibin Zheng and Changyou Xing and Guomin Zhang},
  journal={Computers \& Security},
  year={2023}
}

@article{maeda2021automating,
  title={Automating post-exploitation with deep reinforcement learning},
  author={Maeda, R. and Mimura, M.},
  journal={Computers \& Security},
  
  
  year={2021},
  
}

@inproceedings{gangupantulu2022using,
  title={Using cyber terrain in reinforcement learning for penetration testing},
  author={Gangupantulu, R. and Cody, T. and Park, P. and Rahman, A. and Eisenbeiser, L. and Radke, D. and Clark, R. and Redino, C.},
  booktitle={2022 IEEE International Conference on Omni-layer Intelligent Systems (COINS)},
  
  year={2022},
  organization={IEEE}
}

@article{li2023innes,
  title={{INNES}: An intelligent network penetration testing model based on deep reinforcement learning},
  author={Qianyu Li and Miao Hu and Hua Hao and Min Zhang and Yang Li},
  journal={Applied Intelligence},
  volume={53},
  pages={27110--27127},
  year={2023}
}

@book{Sutton2018,
  author = {Sutton, Richard S. and Barto, Andrew G.},
  edition = {Second},
  publisher = {MIT Press},
  title = {Reinforcement Learning: An Introduction},
  url = {http://incompleteideas.net/book/the-book-2nd.html},
  year = {2018}
}

@article{molina2025training,
  title={Training RL Agents for Multi-Objective Network Defense Tasks},
  author={Molina-Markham, Andres and Robaina, Luis and Steinle, Sean and Trivedi, Akash and Tsui, Derek and Potteiger, Nicholas and Brandt, Lauren and Winder, Ransom and Ridley, Ahmad},
  journal={arXiv preprint arXiv:2505.22531},
  year={2025}
}

@article{anderson2006economics,
  title={The economics of information security},
  author={Anderson, Ross and Moore, Tyler},
  journal={science},
  volume={314},
  number={5799},
  pages={610--613},
  year={2006},
  publisher={American Association for the Advancement of Science}
}

@techreport{anthropicEspionage2025,
    title = {{Disrupting the first reported AI-orchestrated cyber espionage campaign}},
    institution = {Anthropic},
    year = 2025,
    note = {(Online, Accessed 21st January 2026) \url{https://assets.anthropic.com/m/ec212e6566a0d47/original/Disrupting-the-first-reported-AI-orchestrated-cyber-espionage-campaign.pdf}}
}

@article{patterson2024empirical,
  title={Empirical design in reinforcement learning},
  author={Patterson, A. and Neumann, S. and White, M. and White, A.},
  journal={Journal of Machine Learning Research},  
  year={2024}
}

@ARTICLE{Hammar_Intrusion_Prevention_through,
  author={Hammar, Kim and Stadler, Rolf},
  journal={IEEE Transactions on Network and Service Management}, 
  title={Intrusion Prevention Through Optimal Stopping}, 
  year={2022},
  volume={19},
  number={3},
  pages={2333-2348},
  doi={10.1109/TNSM.2022.3176781}}

@misc{Hammar_CSLE,
author = {Hammar, Kim and Stadler, Rolf},
title = {{CSLE: The Cyber Security Learning Environment}},
url = {https://github.com/Kim-Hammar/csle}
}

@misc{msft:cyberbattlesim,
  Author = {Microsoft Defender Research Team.},
  Note = {Created by Christian Seifert, Michael Betser, William Blum, James Bono, Kate Farris, Emily Goren, Justin Grana, Kristian Holsheimer, Brandon Marken, Joshua Neil, Nicole Nichols, Jugal Parikh, Haoran Wei.},
  Publisher = {GitHub},
  Howpublished = {\url{https://github.com/microsoft/cyberbattlesim}},
  Title = {CyberBattleSim},
  Year = {2021}
}

@inproceedings{janisch2023nasimemu,
  title={Nasimemu: Network attack simulator \& emulator for training agents generalizing to novel scenarios},
  author={Janisch, Jarom{\'\i}r and Pevn{\`y}, Tom{\'a}{\v{s}} and Lis{\`y}, Viliam},
  booktitle={European Symposium on Research in Computer Security},
  pages={589--608},
  year={2023},
  organization={Springer}
}

@article{kushner2013real,
  title={The real story of stuxnet},
  author={Kushner, David},
  journal={ieee Spectrum},
  volume={50},
  number={3},
  pages={48--53},
  year={2013},
  publisher={Institute of Electrical and Electronics Engineers, Inc., 345 E. 47 th St. NY~…}
}

@inproceedings{geiger2020analysis,
  title={An analysis of black energy 3, crashoverride, and trisis, three malware approaches targeting operational technology systems},
  author={Geiger, Marcus and Bauer, Jochen and Masuch, Michael and Franke, J{\"o}rg},
  booktitle={2020 25th IEEE International Conference on Emerging Technologies and Factory Automation (ETFA)},
  volume={1},
  pages={1537--1543},
  year={2020},
  organization={IEEE}
}

@misc{karwowski2023goodhartslawreinforcementlearning,
      title={Goodhart's Law in Reinforcement Learning}, 
      author={Jacek Karwowski and Oliver Hayman and Xingjian Bai and Klaus Kiendlhofer and Charlie Griffin and Joar Skalse},
      year={2023},
      eprint={2310.09144},
      archivePrefix={arXiv},
      primaryClass={cs.LG},
      url={https://arxiv.org/abs/2310.09144}, 
}

@misc{ashton2021causalcampbellgoodhartslawreinforcement,
      title={Causal Campbell-Goodhart's law and Reinforcement Learning}, 
      author={Hal Ashton},
      year={2021},
      eprint={2011.01010},
      archivePrefix={arXiv},
      primaryClass={cs.LG},
      url={https://arxiv.org/abs/2011.01010}, 
}

@inproceedings{OeschCyberWheel2024,
author = {Oesch, Sean and Chaulagain, Amul and Weber, Brian and Dixson, Matthew and Sadovnik, Amir and Roberson, Benjamin and Watson, Cory and Austria, Phillipe},
title = {Towards a High Fidelity Training Environment for Autonomous Cyber Defense Agents},
year = {2024},
isbn = {9798400709579},
publisher = {Association for Computing Machinery},
address = {New York, NY, USA},
url = {https://doi.org/10.1145/3675741.3675752},
doi = {10.1145/3675741.3675752},
booktitle = {Proceedings of the 17th Cyber Security Experimentation and Test Workshop},
pages = {91–99},
numpages = {9},
location = {Philadelphia, PA, USA},
series = {CSET '24}
}

@misc{chapman2025r3ace,
  author       = {Chapman, Edward and Hicks, Chris and Mavroudis, Vasilios},
  title        = {r3ace},
  version      = {0.1.0},
  date         = {2025-03-31},
  institution  = {The Alan Turing Institute},
  doi          = {10.5281/zenodo.15147271},
  url          = {https://github.com/alan-turing-institute/r3ace}
}

@misc{dudman2025generalisablecyberdefenceagent,
      title={Towards a Generalisable Cyber Defence Agent for Real-World Computer Networks}, 
      author={Tim Dudman and Martyn Bull},
      year={2025},
      eprint={2511.09114},
      archivePrefix={arXiv},
      primaryClass={cs.LG},
      url={https://arxiv.org/abs/2511.09114}, 
}

@misc{king2025automatedcyberdefensegeneralizable,
      title={Automated Cyber Defense with Generalizable Graph-based Reinforcement Learning Agents}, 
      author={Isaiah J. King and Benjamin Bowman and H. Howie Huang},
      year={2025},
      eprint={2509.16151},
      archivePrefix={arXiv},
      primaryClass={cs.LG},
      url={https://arxiv.org/abs/2509.16151}, 
}

@misc{mern2021autonomousattackmitigationindustrial,
      title={Autonomous Attack Mitigation for Industrial Control Systems}, 
      author={John Mern and Kyle Hatch and Ryan Silva and Cameron Hickert and Tamim Sookoor and Mykel J. Kochenderfer},
      year={2021},
      eprint={2111.02445},
      archivePrefix={arXiv},
      primaryClass={cs.CR},
      url={https://arxiv.org/abs/2111.02445}, 
}

@misc{nyberg2024structuralgeneralizationautonomouscyber,
      title={Structural Generalization in Autonomous Cyber Incident Response with Message-Passing Neural Networks and Reinforcement Learning}, 
      author={Jakob Nyberg and Pontus Johnson},
      year={2024},
      eprint={2407.05775},
      archivePrefix={arXiv},
      primaryClass={cs.AI},
      url={https://arxiv.org/abs/2407.05775}, 
}

@misc{symesthompson2025entitybasedreinforcementlearningautonomous,
      title={Entity-based Reinforcement Learning for Autonomous Cyber Defence}, 
      author={Isaac {Symes Thompson} and Alberto Caron and Chris Hicks and Vasilios Mavroudis},
      year={2025},
      eprint={2410.17647},
      archivePrefix={arXiv},
      primaryClass={cs.LG},
      url={https://arxiv.org/abs/2410.17647}, 
}

@misc{mern2020exchangeableinputrepresentationsreinforcement,
      title={Exchangeable Input Representations for Reinforcement Learning}, 
      author={John Mern and Dorsa Sadigh and Mykel J. Kochenderfer},
      year={2020},
      eprint={2003.09022},
      archivePrefix={arXiv},
      primaryClass={cs.LG},
      url={https://arxiv.org/abs/2003.09022}, 
}

@misc{bates2023rewardshapinghappierautonomous,
      title={Reward Shaping for Happier Autonomous Cyber Security Agents}, 
      author={Elizabeth Bates and Vasilios Mavroudis and Chris Hicks},
      year={2023},
      eprint={2310.13565},
      archivePrefix={arXiv},
      primaryClass={cs.LG},
      url={https://arxiv.org/abs/2310.13565}, 
}

@misc{bates2026rewardsreinforcementlearningcyber,
      title={Beyond Rewards in Reinforcement Learning for Cyber Defence}, 
      author={Elizabeth Bates and Chris Hicks and Vasilios Mavroudis},
      year={2026},
      eprint={2602.04809},
      archivePrefix={arXiv},
      primaryClass={cs.LG},
      url={https://arxiv.org/abs/2602.04809}, 
}

@article{vyas_survey_26,
author = {Vyas, Sanyam and Mavroudis, Vasilios and Burnap, Pete},
title = {{Towards the Deployment of Realistic Autonomous Cyber Network Defence: A Systematic Review}},
year = {2025},
issue_date = {January 2026},
publisher = {Association for Computing Machinery},
volume = {58},
number = {1},
doi = {10.1145/3729213},
journal = {ACM Comput. Surv.},
month = aug,
articleno = {5},
numpages = {36}
}

@article{bates2025less,
  title={Less is more? Rewards in RL for Cyber Defence},
  author={Bates, Elizabeth and Hicks, Chris and Mavroudis, Vasilios},
  journal={arXiv preprint arXiv:2503.03245},
  year={2025}
}

@inproceedings{Agarwal21unstableRL,
author = {Agarwal, Rishabh and Schwarzer, Max and Castro, Pablo Samuel and Courville, Aaron and Bellemare, Marc G.},
title = {Deep reinforcement learning at the edge of the statistical precipice},
year = {2021},
isbn = {9781713845393},
booktitle = {Proceedings of the 35th International Conference on Neural Information Processing Systems},
articleno = {2244},
numpages = {17},
series = {NIPS '21}
}

@INPROCEEDINGS{samaddar25ood,
  author={Samaddar, Ankita and Potteiger, Nicholas and Koutsoukos, Xenofon},
  booktitle={2025 IEEE 4th International Conference on AI in Cybersecurity (ICAIC)}, 
  title={Out-of-Distribution Detection for Neurosymbolic Autonomous Cyber Agents}, 
  year={2025},
  volume={},
  number={},
  pages={1-9},
  doi={10.1109/ICAIC63015.2025.10849024}}

@article{mcfadden2026sok,
  title={SoK: The Pitfalls of Deep Reinforcement Learning for Cybersecurity},
  author={McFadden, Shae and Foley, Myles and Bates, Elizabeth and Tsingenopoulos, Ilias and Vyas, Sanyam and Mavroudis, Vasilios and Hicks, Chris and Pierazzi, Fabio},
  journal={arXiv preprint arXiv:2602.08690},
  year={2026}
}

@misc{emerson2024cyborgenhancedgymdevelopment,
      title={CybORG++: An Enhanced Gym for the Development of Autonomous Cyber Agents}, 
      author={Harry Emerson and Liz Bates and Chris Hicks and Vasilios Mavroudis},
      year={2024},
      eprint={2410.16324},
      archivePrefix={arXiv},
      primaryClass={cs.CR},
      url={https://arxiv.org/abs/2410.16324}, 
}

@inproceedings{cage_1_cyborg_acd_2021,
  author    = {Maxwell Standen and Martin Lucas and David Bowman and Toby Richer and Junae Kim and Damian Marriott},
  title     = {{CybORG}: A Gym for the Development of Autonomous Cyber Agents},
  booktitle = {IJCAI-21 1st International Workshop on Adaptive Cyber Defense},
  year      = {2021}
}

@misc{cage2_kiely2023autonomousagentscyberdefence,
      title={On Autonomous Agents in a Cyber Defence Environment}, 
      author={Mitchell Kiely and David Bowman and Maxwell Standen and Christopher Moir},
      year={2023},
      eprint={2309.07388},
      archivePrefix={arXiv},
      primaryClass={cs.CR},
      url={https://arxiv.org/abs/2309.07388}, 
}

@misc{cage3_code, 
  author = {TTCP CAGE Working Group}, 
  Title = {TTCP CAGE Challenge 3}, 
  Publisher = {GitHub},  
  Howpublished = {\url{https://github.com/cage-challenge/cage-challenge-3}}, 
  Year = {2022} 
}

@article{cage_4_dev_kiely2025cage,
  title={CAGE challenge 4: A scalable multi-agent reinforcement learning gym for autonomous cyber defence},
  author={Kiely, Mitchell and Ahiskali, Metin and Borde, Etienne and Bowman, Benjamin and Bowman, David and Van Bruggen, Dirk and Cowan, KC and Dasgupta, Prithviraj and Devendorf, Erich and Edwards, Ben and others},
  journal={AI Magazine},
  volume={46},
  number={3},
  pages={e70021},
  year={2025},
  publisher={Wiley Online Library}
}

@inproceedings{cage_4_writeup_kiely2025exploring,
  title={Exploring the Efficacy of Multi-Agent Reinforcement Learning for Autonomous Cyber Defence: A CAGE Challenge 4 Perspective},
  author={Kiely, Mitchell and Ahiskali, Metin and Borde, Etienne and Bowman, Benjamin and Bowman, David and van Bruggen, Dirk and Cowan, KC and Dasgupta, Prithviraj and Devendorf, Erich and Edwards, Ben and others},
  booktitle={Proceedings of the AAAI Conference on Artificial Intelligence},
  volume={39},
  number={28},
  pages={28907--28913},
  year={2025}
}

@misc{schwartz2019nasim,
title={NASim: Network Attack Simulator},
author={Schwartz, Jonathon and Kurniawatti, Hanna},
year={2019},
howpublished={\url{https://networkattacksimulator.readthedocs.io/}},
}

@misc{PackerGao_generalisable18,
  Author = {Charles Packer and Katelyn Gao and Jernej Kos and Philipp Kr\"ahenb\"uhl and Vladlen Koltun and Dawn Song},
  Title = {Assessing Generalization in Deep Reinforcement Learning},
  Year = {2018},
  Eprint = {arXiv:1810.12282},
}

@misc{molinamarkham2021networkenvironmentdesignautonomous,
      title={Network Environment Design for Autonomous Cyberdefense}, 
      author={Andres Molina-Markham and Cory Miniter and Becky Powell and Ahmad Ridley},
      year={2021},
      eprint={2103.07583},
      archivePrefix={arXiv},
      primaryClass={cs.CR},
      url={https://arxiv.org/abs/2103.07583}, 
}

@article{Mitre,
  title={Toward a knowledge graph of cybersecurity countermeasures},
  author={Kaloroumakis, Peter and Smith, Michael},
  year={2020}
}

@misc{openC2,
    title={Open Command and Control (OpenC2) Architecture Specification Version 1.0.},
    author= {Duncan Sparrell},
    year = {2022},
    url = {https://docs.oasis-open.org/openc2/oc2arch/v1.0/oc2arch-v1.0.html.}
}

@inproceedings{andrew_Yawning_Titan,
 author = {Andrew, Alex and Spillard, Sam and Collyer, Joshua and Dhir, Neil},
 year = {2022},
 month = {07},
 title = {Developing Optimal Causal Cyber-Defence Agents via Cyber Security Simulation},
 maintitle = {International Confernece on Machine Learning (ICML)},
 booktitle = {Workshop on Machine Learning for Cybersecurity (ML4Cyber)}
}

@inproceedings{Short2025EssentialRoleMSAI,
  author       = {Short, James},
  title        = {The Essential Role of Modelling and Simulation in Helping AI Fight Cyber-Attacks},
  booktitle    = {Force Readiness for Multi-Domain Operations through Modelling and Simulation: NATO Modelling and Simulation Group (MSG) Symposium (MSG-229)},
  series       = {STO Meeting Proceedings},
  number       = {STO-MP-MSG-229},
  year         = {2025},
  note         = {Paper MP-MSG-229-02 (Open Access)},
  publisher    = {NATO Science and Technology Organization (STO)},
  url          = {https://publications.sto.nato.int/publications/STO%20Meeting%20Proceedings/STO-MP-MSG-229/MP-MSG-229-02.pdf}
}

@misc{palmer2024deepreinforcementlearningautonomous,
      title={Deep Reinforcement Learning for Autonomous Cyber Defence: A Survey}, 
      author={Gregory Palmer and Chris Parry and Daniel J. B. Harrold and Chris Willis},
      year={2024},
      eprint={2310.07745},
      archivePrefix={arXiv},
      primaryClass={cs.LG},
      url={https://arxiv.org/abs/2310.07745}, 
}

@misc{MilesEtAl2024RL_ARCD,
  author      = {Miles, Ian and Farmer, Sara and Foster, David and Harrold, Daniel and Palmer, Gregory and Parry, Chris and Willis, Chris and Casassa Mont, Marco and Gralewski, Lisa and Menzies, Ryan and Morarji, Neela and Turkbeyler, Esin and Wilson, Alec and Beard, Alfie and Marques, Pedro and Francis Roscoe, Jonathan and Bailey, Samuel and Cheah, Madeline and Dorn, Mark and Haubrick, Peter and Lacey, Matthew and Rimmer, David and Stone, Jack and Till, Demian and Heartfield, Ryan and Harrison, Andy and Short, James and Wilson, Tom and H, John},
  title       = {Reinforcement Learning for Autonomous Resilient Cyber Defence},
  institution = {Frazer-Nash Consultancy},
  type        = {White paper},
  year        = {2024},
  month       = aug,
  note        = {Presented at Black Hat USA, August 2024},
  url         = {https://www.fnc.co.uk/media/mwcnckij/us-24-milesfarmer-reinforcementlearningforautonomousresilientcyberdefence-wp.pdf}
}

@misc{primaite,
  author       = {{Defence Science and Technology Laboratory UK}},
  title        = {PrimAITE (Primary-level AI Training Environment)},
  version      = {4.0.0},
  date         = {2025-03-18},
  url          = {https://github.com/Autonomous-Resilient-Cyber-Defence/PrimAITE},
  note         = {GitHub repository (tag: v4.0.0). Accessed 2026-02-19}
}

@article{gronauer2022multi,
  title={Multi-agent deep reinforcement learning: a survey},
  author={Gronauer, Sven and Diepold, Klaus},
  journal={Artificial Intelligence Review},
  volume={55},
  number={2},
  pages={895--943},
  year={2022},
  publisher={Springer}
}

@article{huh2023multi,
  title={Multi-agent reinforcement learning: A comprehensive survey},
  author={Huh, Dom and Mohapatra, Prasant},
  journal={arXiv preprint arXiv:2312.10256},
  year={2023}
}

@article{qu2022scalable,
  title={Scalable reinforcement learning for multiagent networked systems},
  author={Qu, Guannan and Wierman, Adam and Li, Na},
  journal={Operations Research},
  volume={70},
  number={6},
  pages={3601--3628},
  year={2022},
  publisher={INFORMS}
}

@article{zhang2024survey,
  title={A survey on self-play methods in reinforcement learning},
  author={Zhang, Ruize and Xu, Zelai and Ma, Chengdong and Yu, Chao and Tu, Wei-Wei and Tang, Wenhao and Huang, Shiyu and Ye, Deheng and Ding, Wenbo and Yang, Yaodong and others},
  journal={arXiv preprint arXiv:2408.01072},
  year={2024}
}

\end{document}